\documentclass[
aip,rsi,
amsmath,amssymb,
longbibliography,
reprint]{revtex4-1}

\usepackage[utf8]{inputenc}
\usepackage[T1]{fontenc}

\usepackage[hidelinks]{hyperref}

\usepackage{graphicx}
\usepackage{dcolumn}
\usepackage{bm}

\usepackage{mathptmx}
\usepackage{etoolbox}

\usepackage{url}

\makeatletter
\def\@email#1#2{%
 \endgroup
 \patchcmd{\titleblock@produce}
  {\frontmatter@RRAPformat}
  {\frontmatter@RRAPformat{\produce@RRAP{*#1\href{mailto:#2}{#2}}}\frontmatter@RRAPformat}
  {}{}
}%
\makeatother
\begin{document}

\title{Measurement and control of a superconducting quantum processor with a fully-integrated radio-frequency system on a chip}

\author{Mats O. Tholén}
\thanks{contributed equally to this work}
\affiliation{Nanostructure Physics, KTH Royal Institute of Technology, 106 91 Stockholm, Sweden}
\affiliation{Intermodulation Products AB, 823 93 Segersta, Sweden}

\author{Riccardo Borgani}
\thanks{contributed equally to this work}
\affiliation{Nanostructure Physics, KTH Royal Institute of Technology, 106 91 Stockholm, Sweden}
\affiliation{Intermodulation Products AB, 823 93 Segersta, Sweden}

\author{Giuseppe Ruggero Di Carlo}
\affiliation{Nanostructure Physics, KTH Royal Institute of Technology, 106 91 Stockholm, Sweden}

\author{Andreas Bengtsson}
\author{Christian Kri\v{z}an}
\author{Marina Kudra}
\author{Giovanna Tancredi}
\author{Jonas Bylander}
\author{Per Delsing}
\author{Simone Gasparinetti}
\affiliation{Department of Microtechnology and Nanoscience, Chalmers University of Technology, 412 96 Gothenburg, Sweden}

\author{David B. Haviland}
\affiliation{Nanostructure Physics, KTH Royal Institute of Technology, 106 91 Stockholm, Sweden}
\email[The author to whom correspondence may be addressed: ]{haviland@kth.se}

\date{2022-05-28}

\begin{abstract}

We describe a digital microwave platform called Presto, designed for measurement and control of multiple quantum bits (qubits) and based on the third-generation radio-frequency system on a chip.
Presto uses direct digital synthesis to create signals up to 9~GHz on 16 synchronous output ports, while synchronously analyzing response on 16 input ports.
Presto has 16 DC-bias outputs, 4 inputs and 4 outputs for digital triggers or markers, and two continuous-wave outputs for synthesizing frequencies up to 15~GHz.
Scaling to a large number of qubits is enabled through deterministic synchronization of multiple Presto units.
A Python application programming interface configures a firmware for synthesis and analysis of pulses, coordinated by an event sequencer.
The analysis integrates template matching (matched filtering) and low-latency (184 -- 254~ns) feedback to enable a wide range of multi-qubit experiments.
We demonstrate Presto's capabilities with experiments on a sample consisting of two superconducting qubits connected via a flux-tunable coupler.
We show single-shot readout and active reset of a single qubit; randomized benchmarking of single-qubit gates showing 99.972\% fidelity, limited by the coherence time of the qubit; and calibration of a two-qubit iSWAP gate.

\end{abstract}

\maketitle

\section{Introduction}

Quantum technology is presently enjoying rapid expansion, driven in part by the promise of speed-up in the processing of information \cite{harrow_quantum_2017}.
Several large-scale projects are underway to develop a fully-programmable quantum-information processing machine, based on a variety of different quantum hardware \cite{ladd_quantum_2010}.
Common to all these developments is the need for a classical electronic control system that is flexible, and possible to scale up in a cost-efficient manner.
In fact, one of the main impediments to scaling up a quantum computer is the cost and complexity of the classical control system \cite{riste_microwave_2020, asaad_independent_2016, mcdermott_quantumclassical_2018}.
Here we address this issue through the development of a fully-programmable digital microwave control platform for a superconducting quantum processor.
Our direct-digital approach to synthesize and readout signals to and from the quantum chip differs from traditional qubit-control systems that use analog mixers to up- and down-convert microwave signals \cite{KrantzP2019Aqeg}.

Contemporaneous with the expansion of quantum technology, high-speed digital circuits have undergone significant advancement, driven by the deployment of the fifth-generation technology standard for broadband cellular networks (5G) and the transition to software-defined radio (SDR).
5G and SDR have stringent hardware and signal-processing requirements for receiving and transmitting signals with phased arrays of antennas, requiring many phase-coherent radio-frequency (RF) channels, each with high bandwidth, low distortion and low noise.
A tight integration of RF data converters and digital signal processing (DSP) is desired to cope with high data rates while keeping the infrastructure cheap, scalable and power efficient.
Such requirements should sound familiar to quantum technologists.
Indeed, much is to be gained by taking advantage of 5G and SDR development, adapting this new digital hardware to quantum-technology applications.

A notable advancement in this context is the Zynq UltraScale+ RFSoC, a radio-frequency system on a chip integrating RF data converters, many cores of central-processing units (CPU) and a large field-programmable gate array (FPGA), all on a single silicon chip \cite{rfsoc}.
Evaluation modules of the first generation of RFSoC chips have been used to demonstrate their applicability to quantum computation \cite{Gebauer2020-1, Gebauer2020-2, stefanazzi_qick_2022, park2021icarusq, Singhal2022}.
FPGA devices have been used quite extensively in scientific experiments \cite{carminati2021impact} and there are many examples of their use for readout and control of superconducting qubits \cite{walter2017singleshot, Heinsoo2018, Salathe2018, xu2021qubic}.
The RFSoC platform differs significantly from this previous use of FPGAs because the on-chip integration of high-speed data converters enables direct digital synthesis (DDS) of RF waveforms.

Especially the third-generation RFSoC allows for the generation of output signals and real-time analysis of input signals up to 9~GHz.
With proper care \cite{vandijk2020ddsbased, Ball2016roleofmaster}, DDS can reliably generate the microwave signals to control a quantum processor, and synchronously analyze signals from the readout of qubits, with high fidelity and in a compact and scalable format.
The RFSoC eliminates external local oscillators and analog IQ mixers, expensive components in traditional control systems, with imperfections that necessitate calibration and compensation \cite{jolin2020calibration}.

In this paper we describe Presto, a digital microwave platform for signal generation, acquisition and processing based on the third generation of RFSoC.
Presto is a multi-purpose platform with a wide variety of features that are programmable from a Python application programming interface (API).
We use Presto to characterize a quantum processor, implementing high-fidelity single-qubit gates and achieving high-fidelity single-shot readout.
We implement a conditional reset operation \cite{riste_feedback_2012} on a single qubit, essentially cooling it with Presto's low-latency feedback engine.
We perform randomized benchmarking on single qubit gates using Presto's internal averaging capabilities.
We also use multiple synchronized channels of Presto to tune up and implement a two-qubit gate, with the readout of both qubits multiplexed on the same channel.
In all experiments both measurement and control is performed by Presto alone.
No additional electronic instruments were required.

This paper is organized as follows:
In Sec.~\ref{sec:platform} we describe the Presto platform based on the RFSoC.
Sections~\ref{sec:digital} and~\ref{sec:analog} describe in detail the architecture of Presto's digital and analog components, respectively, motivating the design choices for the readout and control of superconducting qubits.
In Sec.~\ref{sec:scaling} we discuss how multiple Presto units are synchronized to scale up toward controlling hundreds of qubits.
In Sec.~\ref{sec:exp} we demonstrate the use of Presto in a series of experiments on a two-qubit quantum processor.
Appendixes~\ref{apx:cw} and~\ref{apx:match} provide additional details on continuous-wave capabilities of Presto and on the use of template matching for implementing optimal readout, respectively.
Appendix~\ref{apx:qutrit} discusses the implementation of a feedback scheme for active qutrit reset and Appendix~\ref{apx:setup} gives the details our qubit-measurement setup.
Finally, Appendix~\ref{apx:abbrev} lists the abbreviations used in this manuscript together with their explanation.

\section{Architecture}

\subsection{RFSoC platform}
\label{sec:platform}

\begin{figure}
  \centering
  \includegraphics{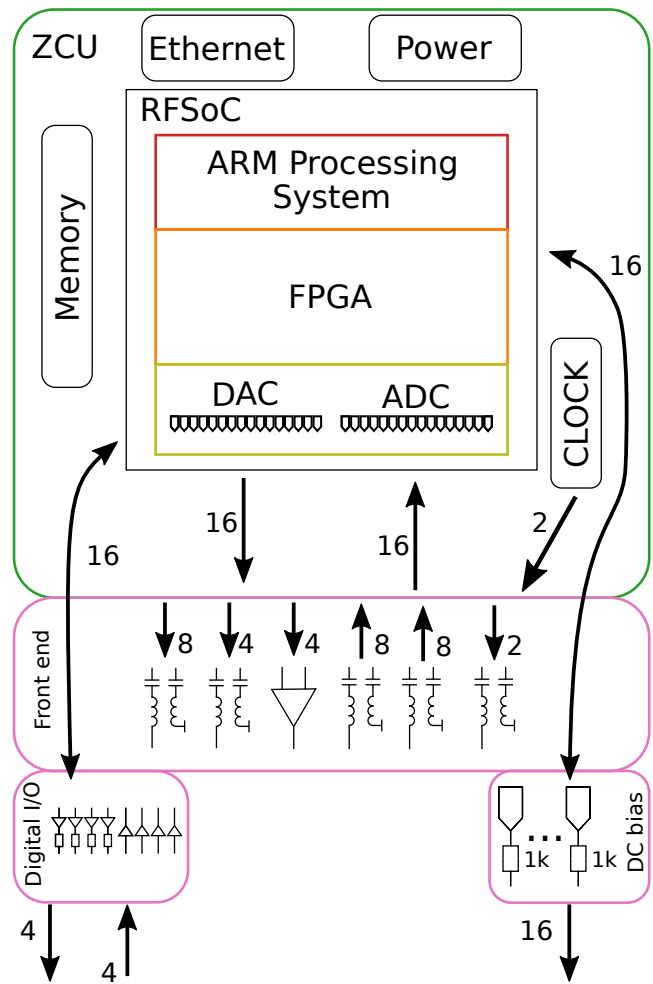}
  \caption{
    Overview of Presto's circuit boards and subsystems.
    The ZCU208 or ZCU216 development board from Xilinx (green) houses the RFSoC chip and other components such as memory, clock synthesizers and network controllers.
    Three custom boards (pink) provide additional functionality:
    two boards convert signals from the RFSoC to a more useful format,
    and a third board provides 16 DC-biasing channels.
  }
  \label{fig:ArchitectureBoards}
\end{figure}

The core of the Presto platform is the third generation of Zynq UltraScale+ RFSoC by Xilinx \cite{rfsoc}.
These highly-integrated chips feature a full ARM processing subsystem, a large FPGA and numerous RF data converters.
The current versions of Presto do not use the bare chips directly, but rather build on the development boards \cite{zcus} ZCU208 or ZCU216, depending on the number of RF channels needed:
the former board features 8 input and 8 output channels,
while the latter features 16 of each.
The symmetric input/output design is optimal for SDR applications where phased arrays of antennas used for beamforming are typically symmetrical on the transmitting and receiving end.
The symmetric configuration is however not optimal for quantum-computing applications where several output channels are required for each input channel.
The current architecture of superconducting quantum processors favours frequency-multiplexed qubit readout, but individual microwave lines are required to control the qubits.

In principle, these development boards contain everything needed to operate the RFSoC.
In practice, the analog and digital signals are not exposed in a suitable manner for interfacing to quantum circuits.
To address this problem, we developed daughter boards (Fig.~\ref{fig:ArchitectureBoards}) to convert the RF signals and digital triggers and markers to a more useful form.
An additional board adds the capability of supplying DC bias to flux-tunable qubit couplers or parametric amplifiers.
We also exposed two high-frequency drive ports (up to 15~GHz) for pumping parametric amplifiers.
The daughter boards are described in detail in Sec.~\ref{sec:analog}.

The RF channels are provided by 14-bit analog-to-digital (ADC) and digital-to-analog (DAC) RF converters that are tightly integrated on-chip with the programmable logic.
The ADCs and DACs are designed to directly measure and synthesize microwave signals in higher Nyquist zones.
The bandwidth at maximum output power is specified \cite{rfsocDSoverview, rfsocDSacdc} to 6~GHz.
However, our measurements show that the RFSoC generates significant signal power well above 8~GHz with our front-end board.
As shown in detail in Sec.~\ref{sec:exp} and Fig.~\ref{fig:outrng}, Presto comfortably accesses the 4-to-8~GHz band typically used in cryogenic microwave setups.

\subsection{Digital design}
\label{sec:digital}

Direct access to all converters from the FPGA allows for real-time digital synthesis and analysis of intricate waveforms that are fully programmable, enabling a wide variety of measurements.
Programming the FPGA gives the most flexibility, but such low-level programming is a tedious and difficult task that requires competence and expertise not typically available in a quantum-physics laboratory.
For this reason we focused on designing different modes of operation, each with a generous set of measurement-configuration parameters that are controlled from a Python API \cite{presto-api}.
These modes of operation are implemented as different firmware that configures Presto at run time.
Presently two general-purpose modes of operation are available: one for continuous-wave and one for pulsed experiments.

The continuous-wave mode of operation has been used to investigate multipartite entanglement of microwave modes in surface-acoustic wave resonators \cite{andersson2022squeezing} and Josephson parametric amplifiers \cite{jolin2021multipartite}.
Another potential application of this firmware is frequency-multiplexed sensor arrays \cite{echternach2018,mazin_position_2006}, where a large number of superconducting microwave resonators on one transmission line are pumped at different frequencies while continuously monitoring changes in their response amplitude and phase.
We provide some detail on continuous-wave operation in Appendix~\ref{apx:cw}.

In this manuscript we focus mainly on the pulsed mode of operation.
An early version of this firmware has been used to create non-classical states of the microwave field in a superconducting 3D cavity \cite{kudra_robust_2022}.

\subsubsection{Digital up- and down-conversion}
\label{sec:updownconv}

Digital up- and down-conversion are well-defined numerical operations and the accuracy of these DSP techniques is limited only by the number of bits \cite{vandijk2020ddsbased}.
Digital frequency conversion does not suffer from the issues that affect its analog counterpart, such as local-oscillator leakage, amplitude and phase imbalance, drift or frequency-dependent nonlinearity.

Presto outputs pulses and analyzes response using DSP on the FPGA.
Each sample in the data stream is a complex number, encoding a pair of in-phase (I) and quadrature (Q) components.
The FPGA is clocked at 500~MHz and operates on two samples in parallel, resulting in a complex data stream with a sampling rate of 1~GS/s, on each of the 16 DAC and ADC channels.

On the pulse generation side, 16-bit signals generated by the FPGA are interpolated to a user-selected sampling rate (up to 10~GS/s) and up-converted with a fully-digital IQ mixer and a numerically-controlled oscillator (NCO).
The NCOs run at the full sampling rate of the DAC and have 48-bit frequency resolution and 18-bit phase resolution.
The data is then truncated to 14 bits before being output by the DAC.
With this scheme the DAC can output signals in a $\pm$500~MHz band centered around the NCO frequency, programmable between 0 and 10~GHz with less than $40~\mu$Hz resolution.

On the measurement side, signals are sampled by the ADC at the user-selected sampling rate (up to 5~GS/s) with 14-bit resolution and then extended to 16 bits to accommodate increased accuracy in subsequent filtering stages.
Data is then down-converted to baseband with a fully-digital IQ mixer and NCO.
At baseband the signals are decimated to 1~GS/s and analyzed as two complex samples in parallel by the FPGA.

Presto has one independently programmable digital IQ mixer and NCO for every input and output channel.
If not all physical ports are needed, the digital mixers and NCOs from neighbouring channels can be combined and used on a single physical port, further extending the generation and acquisition of complex signals.

Digital up- and down-conversion can be disabled, in which case the DSP operates with a data stream of real numbers at a rate of 2~GS/s.
In this direct-output mode of operation Presto performs a more traditional form of DSP, where signals are generated and acquired directly at baseband, for example to drive parametric coupling of two qubits or for up- and down-conversion with external analog IQ mixers.

\subsubsection{Event sequencer}

\begin{figure}
  \centering
  \includegraphics{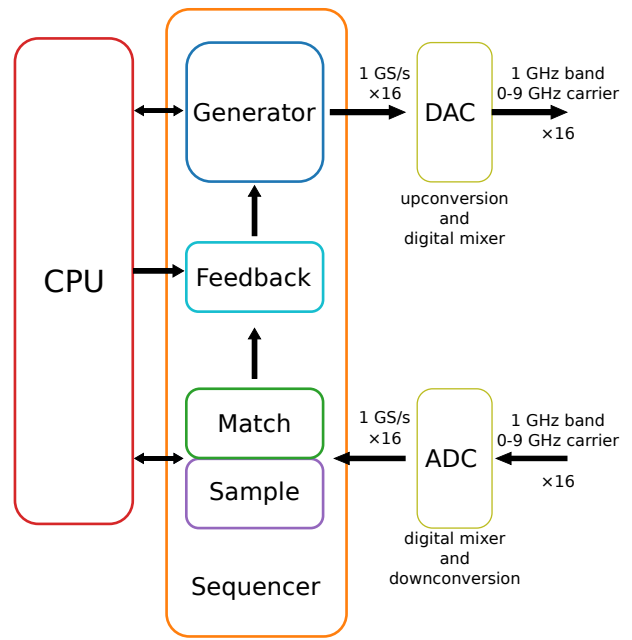}
  \caption{
    High-level functional schematics of the pulsed mode of operation.
    At run time the event sequencer controls the signal generators for the outputs
    and the sampling and template-matching units for the inputs.
    A feedback unit connects the outcome of input-signal analysis to the pulse generation.
  }
  \label{fig:pulsedtop}
\end{figure}

The architecture of the pulsed mode of operation is shown in Fig.~\ref{fig:pulsedtop}.
The primary design goal was gate-based quantum computing where accurate timing of events, such as the application of control pulses and the analysis of readout pulses, is of primary importance.
Presto controls these events with a sequence generator, or sequencer, implemented partially in the FPGA and partially as a `bare-metal' application (\textit{i.e.} running without an operating system) on a real-time processing unit (RPU).
The RPU application is written in the Rust programming language, allowing for the generation of very complex measurement sequences that would be hard to implement in a generic and efficient way on the FPGA.
To maintain performance and real-time control of the sequence, we implemented time-controlled first-in-first-out (FIFO) buffers.

An experiment sequence is first specified at the level of the Python API \cite{presto-api}, and then translated into a sequence of events scheduled on a time line with a 2~ns grid (the inverse of the FPGA clock frequency).
While the sequencer time grid is 2~ns, finer resolution can be achieved by adjusting waveform templates (see below).
Ultimately the timing accuracy in the execution of events is limited by the jitter of the master clock, which we measure to be much less than 1~ps.

Before the experiment is started, all user-defined data and parameters are uploaded into Presto, together with a list of timed events that describe the experimental sequence.
Many different kinds of events can be scheduled, each event tied to a feature of signal generation, acquisition, processing and feedback.
We describe each feature in detail in the next sections.

Once the sequence is started and the experiment begins, the sequencer controls all signal generation, data acquisition and feedback with precise timing.
The instrument performs the whole measurement without interaction or communication with the controlling computer.
When the measurement is finished the acquired data that is stored in Presto is downloaded to the controlling computer for further analysis.
A single measurement sequence can last up to $\approx$6.5 days, or $2^{48}$ clock cycles.
A sequence can be repeated many times to internally perform averaging and/or parameter sweeps, for a virtually unlimited total run time ($\approx$1.2 thousand years, or $2^{64}$ clock cycles).

\subsubsection{Signal generation}

\begin{figure}
  \centering
  \includegraphics{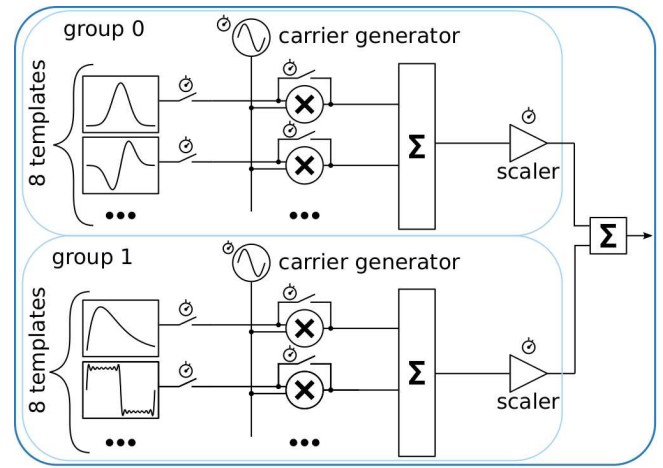}
  \caption{The signal generator for the outputs is built on templates (arbitrary waveforms) and carrier generators.
  Templates are uploaded before the experiment.
  During the experiment, the event sequencer controls template-output timing,
  frequency and phase of the carrier generators and scale factors.
  The resources shown in this figure are replicated on each of the output channels.
  }
  \label{fig:DigitalOutput}
\end{figure}

Figure~\ref{fig:DigitalOutput} shows a functional diagram of signal generation.
Each output channel has 16 templates available for arbitrary waveforms, each storing up to 1022~ns of data, \textit{i.e.} 1022 IQ pairs at 1~GS/s, with 16-bit resolution.
Templates are designed by the user and uploaded to Presto via the API, prior to execution.
The output from each template is controlled individually by the sequencer that can concatenate, superimpose and play templates in a loop, constructing complex waveforms in real time.

The 16 templates are divided into two groups of 8 within each channel, with each group having a dedicated carrier generator and a variable gain (scalar multiplier).
Splitting the available templates into two groups with independent carrier generators and independent scaling allows Presto to efficiently synthesize displacements in the IQ plane, a useful feature for quantum state tomography of bosonic modes \cite{kudra_robust_2022}.

The carrier generators have 40-bit frequency resolution (0.5~mHz) and independent phase offsets for both I and Q in the data stream, with 40-bit resolution (6~prad).
The generators are also controlled by the sequencer which can retrieve parameters from look-up tables with 512 entries at 2~ns intervals, to change the frequency and phase of the carrier, in real time.

As shown in Fig.~\ref{fig:DigitalOutput}, templates are either output as raw data, bypassing the carrier generator, or multiplied by the carrier to create envelopes that modulate the carrier tone.
Because the carrier generator and templates use the same sampling rate, the exact same output waveform could be achieved either as a raw template or by multiplying a carrier with an envelope template.
The main purpose of the carrier generator is therefore not up-conversion, but rather efficient implementation of \textit{e.g.} frequency sweeps, without having to upload many templates that differ only in their carrier frequency.
Frequency up-conversion is performed with the dedicated digital IQ mixer and NCO described in Sec.~\ref{sec:updownconv} and Fig.~\ref{fig:pulsedtop}.

Data from templates in each group (either raw or as envelopes) are summed together and fed to a signed scalar multiplier with 17-bit resolution.
This scaling is controlled by the sequencer, with gain values taken from a user-programmable look-up table with 512 entries.
The two scaled outputs are finally summed together and sent to the digital up-conversion chain described in Sec.~\ref{sec:updownconv}.

\subsubsection{Signal acquisition}
\label{sec:acquisition}

\begin{figure}
  \centering
  \includegraphics{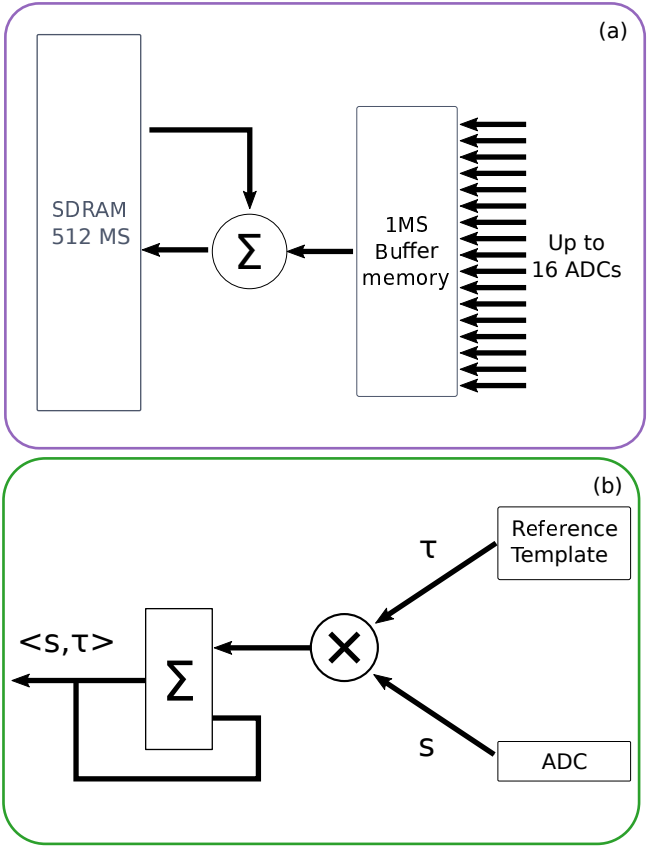}
  \caption{
    The signal inputs have two parallel paths that operate independently:
    (a) for sampling and storage, and (b) for template matching.
    Sampling windows are controlled by the sequencer.
    Samples are initially stored in a high-bandwidth buffer and then averaged and stored in a larger but slower memory.
    Template matching multiplies the incoming samples with a reference template in real time, returning the sum at the end of the matching window.
  }
  \label{fig:sampling}
\end{figure}

There are two methods of analyzing signals sampled by the ADC after the digital down-conversion chain described in Sec.~\ref{sec:updownconv}: either with a store operation or with a template-matching operation.
The two operations can be performed in parallel on the same data stream.

The store operation [Fig.~\ref{fig:sampling}(a)] is similar to boxcar averaging, but with the capability to interleave averaging while stepping parameters.
A high-bandwidth buffer is shared among all the input channels and it is capable of simultaneously receiving up to 16 data streams of complex values, each at 1~GS/s.
The entire buffer memory, up to 524~$\mu$s of data or $2^{19}$ IQ pairs, is available for storage regardless of how many channels are streaming.

Data in this buffer is then transferred at a rate of 1~GS/s to a larger synchronous dynamic random-access memory (SDRAM) capable of storing up to $2^{29}$ samples with 32-bit precision (2~GB), equivalent to $268$~ms of sampled data.
Each transfer can target a particular address in memory, where new data is summed to existing data in the SDRAM.
Together with the stepping of parameters by the sequencer, this architecture enables interleaved averaging of measurement results, to efficiently reject slow drift arising from temperature or gain variations in the measurement system.

The second method of analyzing signals is template matching [Fig.~\ref{fig:sampling}(b)], where the incoming signal $s$ (a complex-valued data stream) is multiplied with a reference template $\tau$ (complex-valued arbitrary waveform) and accumulated over the template length.
The result is the overlap sum of the measured signal with the template,
\begin{equation}
  \left< s, \tau \right> \stackrel{\text{def}}{=} \Re \left\{ \sum \tau^* s \right\}
\end{equation}
giving a single number that quantifies how well the signal matches the template.
Template matching can be used to perform standard IQ demodulation:
Loading one template with $\cos(\omega t)$ and another with $-\sin(\omega t)$ produces the I and Q components of the signal at the frequency $\omega$.
Another use of template matching is to efficiently reduce data.
Template matching is done in real time and does not rely on data stored in the buffer memory.

Template matching with a known reference trace implements a so-called matched filter, an optimal linear filter maximizing signal-to-noise ratio.
One application of matched filtering is quantum state discrimination for the readout of superconducting qubits.
Here two reference templates are calibrated from the measured response of the readout resonator, one when the qubit is in the ground and the other when the qubit is in the excited state.
Performing template matching then results in two numbers, describing how well the input matches each of the two references.
The extension to readout of a three-level qutrit, and beyond, is realized by matching to more templates as discussed in Appendix~\ref{apx:qutrit}.
There are a total of 128 template-matching units for simultaneous discrimination of multiple states, on multiple channels.

\subsubsection{Feedback}
\label{sec:feedback}

\begin{figure}
  \centering
  \includegraphics{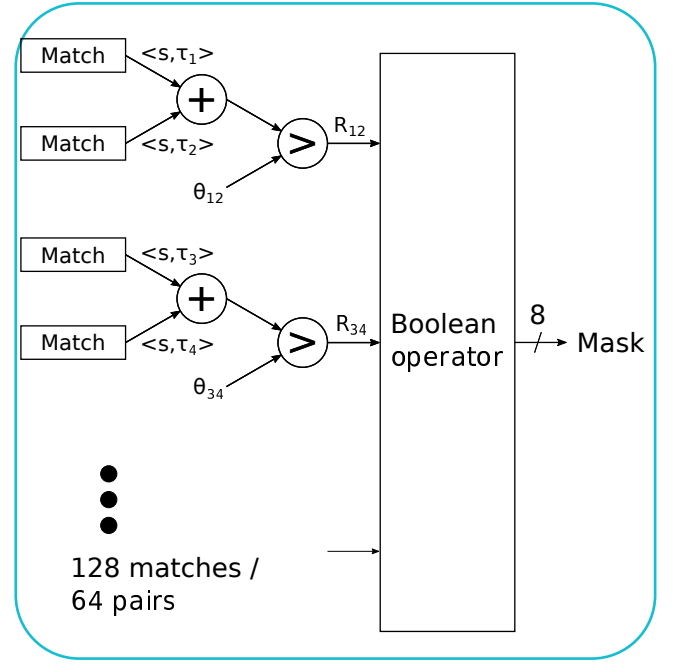}
  \caption{The feedback system does pairwise addition on template-matching outputs, comparing the result to a threshold.
  The Boolean results of the comparisons are subjected to a logical operator to create a mask that gates output templates.}
  \label{fig:feedback}
\end{figure}

Figure~\ref{fig:feedback} shows a schematic of the feedback.
The outcome of the 128 template-matching units, $\left< s, \tau_i \right>$ with $i\in [0,127]$, are used to enable and disable output pulses with low latency.
The outcomes are first added pairwise, $\left< s, \tau_i \right> + \left< s, \tau_{i+1} \right>$, and then compared to a user-programmable threshold $\theta_{i,i+1}$.
If the outcome is greater than the threshold, Boolean value True is generated, otherwise False.
These 64 Boolean values are fed to a configurable operator to generate an 8-bit mask.
The output templates can be gated with the bits of this mask.

As an example, we describe the use of this feedback feature for active reset of a qubit.
We begin by calibrating reference templates $\tau_\mathrm{g}$ and $\tau_\mathrm{e}$ for the readout of the qubit in its ground and excited state, respectively (see Appendix~\ref{apx:match}).
We upload $\tau_\mathrm{e}$ and $-\tau_\mathrm{g}$ into two of the 128 available slots in the template-matching engine.
We program a threshold $\theta_{eg} = (\lVert \tau_\mathrm{e} \rVert^2 - \lVert \tau_\mathrm{g} \rVert^2)/2$,
where $\lVert \tau \rVert^2 = \left< \tau, \tau \right>$.
We also program an output pulse $\pi_\mathrm{eg}$ to flip the qubit from the excited to the ground state, \textit{i.e.} a $\pi$ pulse.
Finally, we gate the output pulse $\pi_\mathrm{eg}$ using the first bit of the feedback mask.
We realize this with a trivial Boolean operator that simply sets this bit to 1, if the sum of the first template-matching pair is greater than the threshold.

The user schedules the template matches on both $\tau_\mathrm{e}$ and $-\tau_\mathrm{g}$ in the sequencer, with the output of the conditional $\pi_\mathrm{eg}$ right after the end of the matching window.
$\pi_\mathrm{eg}$ will be output only if the acquired signal trace matches $\tau_\mathrm{e}$ better than $\tau_\mathrm{g}$, \textit{i.e.} only if the qubit is most likely in its excited state.
We show the results of experiments with this active-reset technique in Sec.~\ref{sec:reset}.
We discuss the implementation of a more advanced feedback algorithm, the reset of a quantum three-level system, or qutrit, in Appendix~\ref{apx:qutrit}.

The total latency of the feedback engine has many contributions: propagation delay through the experimental setup, latency in the RF data converters, and processing latency in the programmable logic.
The latter introduces 5 clock cycles of latency, equivalent to 10~ns.
Propagation delay is dependent only on the experimental setup external to Presto, and is typically about 5~ns per meter of coaxial cable.
The total latency is dominated by the contribution from the high-speed ADCs and DACs.
This contribution is intrinsic to the RFSoC platform, and varies depending on the features enabled in the converters and on the chosen sampling rate.
Over the different converter configurations that are most useful, we measure a combined total round-trip latency between 184~ns and 254~ns.

The round-trip latency can play an important role in the design of experiments and algorithms that involve feedback, such as the active reset of a qubit described above.
In such experiments the latency $\lambda$ imposes a minimum reaction time between the measurement of the qubit in its excited state and the application of a $\pi$ pulse to reset it to its ground state.
During this ``dead time'', the qubit could spontaneously decay to the ground state with probability $1 - \mathrm{e}^{- \lambda / T_1}$, and thus the $\pi$ pulse would erroneously excite the qubit again.
Because $\lambda$ is dominated by the intrinsic latency in the RFSoC platform, lowering the error rate amounts to improving the quality of the qubit by increasing its relaxation time $T_1$.

\subsection{Analog design}
\label{sec:analog}

\begin{figure}
  \centering
  \includegraphics{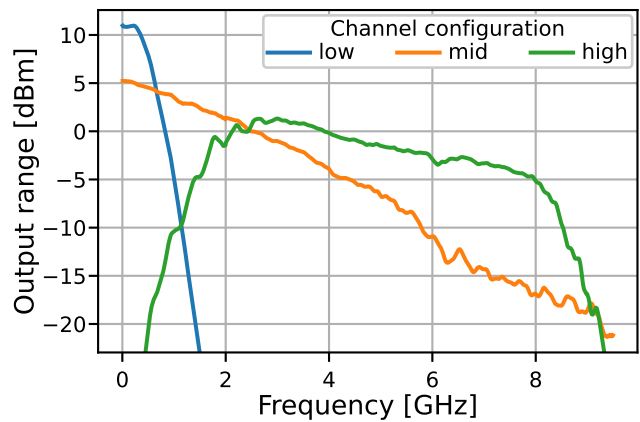}
  \caption{Full-scale output range of the RF channels with the front-end card as function of frequency.
  The curves show the performance for the high-frequency (green), mid-frequency (orange) and low-frequency (blue) configurations.
  }
  \label{fig:outrng}
\end{figure}

The signals leaving and entering the RFSoC development boards are not well suited for interfacing with common experimental setups and auxiliary instruments.
The RF signals are differential with 100$\Omega$ impedance, centered around a rather large common-mode voltage.
Digital signals are 1.8~V low-voltage transistor-transistor logic (LVTTL), which is rather low for interfacing to a wide variety of instruments.
Moreover, both RF and digital signals are connected directly to the FPGA pins, increasing the risk that an accidental static discharge could permanently damage the most delicate and expensive component of the platform.

To address these issues we designed three printed circuit boards (PCB) to transform signals to and from the development board.
The first PCB is a front-end board which connects directly to the development board through two high-density, high-bandwidth, board-to-board connectors.
The front-end board contains the analog circuits for all RF signals and it fans out power and digital signals to the two other PCBs: one for digital triggers/markers, and one for DC biasing.

The front-end board is designed for versatility, with superconducting qubit experiments in mind.
The design makes all 16 output and 16 input channels of the ZCU216 available to the user.
The 16 outputs come in three variants: 8 high frequency, 4 mid frequency and 4 low frequency.
The high and mid frequency variants use AC-coupling capacitors to reject the common-mode voltage and a balun transformer to convert from 100$\Omega$ differential to 50$\Omega$ single-ended signals.
The only difference between high and mid frequency range is the choice of balun transformer:
the high frequency outputs use a 3000-8000~MHz transformer (Mini-Circuits NCS2-83+),
and the mid frequency outputs use a 10-3000~MHz transformer (Mini-Circuits TCM2-33WX+).
The low frequency outputs use a differential to single-ended amplifier (Texas Instruments THS3217), allowing these channels to be DC coupled and providing more output power, at the cost of a much lower bandwidth of 500~MHz.

Figure~\ref{fig:outrng} shows the frequency response of the three different variants.
With this front-end configuration, one Presto unit can \textit{e.g.} drive one readout line and 7 qubit control lines using the 8 high-frequency channels, with the 8 mid and low frequency channels driving the couplers.
On the input side all 16 inputs are AC coupled, half of them use the high frequency and half of them use the mid frequency configuration.

The RFSoC development boards also feature two high-frequency clock outputs, user programmable up to 15~GHz.
These signals are also differential with $100~\Omega$ impedance, so we transform them with AC-coupling capacitors and a 12-18~GHz balun transformer (Mini-Circuits NCR2-183+) before exposing them to the front panel.
These two outputs can be configured independently in frequency and output power, and they are locked in frequency and phase to all of Presto's inputs and outputs.
They are, however, limited to continuous-wave operation at a single frequency.
For superconducting quantum circuits, these outputs are useful for pumping parametric amplifiers such as Josephson parametric amplifiers (JPA) \cite{Aumentado2020} and travelling-wave parametric amplifiers (TWPA) \cite{Planat2020}.

The second PCB processes digital signals to be used as triggers and markers.
It exposes 4 inputs and 4 outputs, each with a bus transceiver that makes the signals 3.3~V compatible and a resistive network to maintain a well defined voltage on open inputs and to provide some protection against static discharge.
Outputs have 50$\Omega$ source impedance and each uses two transceivers in parallel to provide enough current to drive a 50$\Omega$ load.
The inputs have an impedance of 10~k$\Omega$ so that they can be driven by low-power CMOS sources.
The outputs can trigger or mask other instruments and they are controlled by the event sequencer, time-aligned with all other events in the experimental sequence.
For example, a digital marker can gate an external pump to a parametric amplifier, so that it is active only while reading out a resonator, thus reducing noise propagating from the amplifier to the qubits.
The digital inputs are also available to the programmable logic and the event sequencer.

The third PCB provides DC-biasing capabilities.
It has 16 independent voltage outputs with 16-bit resolution, $1~k\Omega$ source impedance, and five user-selected ranges: 3.3~V, 6.6~V, $\pm$3.3~V, $\pm$6.6~V and $\pm$10~V.
These DC voltages are regulated by an on-board reference with $<$125~ppm accuracy and their noise density at 1~kHz is 60~$\mathrm{nV}/\sqrt{\mathrm{Hz}}$, or 1.7~$\mathrm{\mu V}$~rms integrated over the band 0.1~-~10~Hz.
The outputs are controlled in real time during an experiment by the event sequencer.

\subsection{Synchronization and scaling}
\label{sec:scaling}

Multi-channel synchronization and timing are critical features for quantum experiments.
All clocks for the programmable logic, the RF converters and the high-frequency continuous outputs, are generated from a single master clock in the RFSoC.
Extra care is put into ensuring that Presto's entire clock-generation chain has a single, deterministic latency.
We achieved this synchronization with an appropriate configuration of the on-board phase-locked loops (PLLs), and through the use of an RFSoC feature called multi-tile synchronization (MTS).
MTS ensures that the sampling of all input and output channels, which might run at different rates, are frequency and phase aligned.
We further augmented MTS with a calibration to ensure that the alignment is reproducible across reconfiguration of the programmable logic.

With this level of synchronization across all 16 channels, we measure the phase noise of Presto to be -115.26~dBc/Hz at 10~kHz offset from a 3~GHz carrier, and the rms phase jitter to be 105~fs integrated between 12~kHz and 20~MHz.
This phase noise is equivalent to -124.80~dBc/Hz on a 1~GHz carrier, or -109.24~dBc/Hz on a 6~GHz carrier.
The jitter between any two output channels is less than 1~ps, limited by the measuring oscilloscope.

The maximum number of outputs on a single Presto unit is 16, however, much higher channel count is required for the current architecture of quantum processors as they scale up the number of qubits and couplers \cite{krinner2019,Lecocq2021}.
To meet this challenge we support the synchronization of multiple Presto units with a dedicated device which we call Metronomo.
Metronomo is a clock source that exposes 7 pairs of device and SYSREF clocks.
The device clocks synchronize different Presto units with a programmable reference frequency.
Typically 10~MHz or 100~MHz clock reference is used, but they can be as high as 3~GHz.
The SYSREF signals are single synchronization pulses, precisely aligned to the device clock outputs.
These serve as timing references to provide low skew and jitter between different Presto units.
With Metronomo we observe that the jitter between two different Presto units is indistinguishable from the jitter between outputs on the same unit.

One can synchronize up to 7 Presto units with one Metronomo at the center of a star configuration, for a total 112 synchronized output channels.
If more outputs are need, multiple Metronomos are designed to be connected in a cascade: one Metronomo can synchronize 7 other Metronomos, which in turn synchronize 7 Presto units each for a total of 784 output channels.
While Metronomo is designed to support such a configuration, the performance of the cascaded setup has not yet been tested.

\section{Experiments}
\label{sec:exp}

\subsection{Qubit sample and measurement configuration}

\begin{table}
  \centering
  \begin{tabular}{p{0.3\columnwidth} p{0.3\columnwidth} p{0.3\columnwidth}}
    \hline
    \hline
    Parameter & Qubit 1 & Qubit 2 \\
    \hline
    $\omega_\mathrm{R} / 2\pi$ & 6.17 GHz & 6.03 GHz \\
    $\kappa / 2\pi$ & 615 kHz & 455 kHz \\
    $\omega_{01} / 2\pi$ & 3.56 GHz & 4.09 GHz \\
    $\alpha / 2\pi$ & -240 MHz & -231 MHz \\
    $\chi / 2\pi$ & -155 kHz & -302 kHz \\
    $g / 2\pi$ & 69.3 MHz & 74.3 MHz \\
    $\chi / \kappa$ & -0.25 & -0.66 \\
    $\Gamma^{-1}_{Purcell}$ & 370 $\mu$s & 240 $\mu$s \\
    $T_1$ & 46 $\pm$ 7 $\mu$s & 34 $\pm$ 9 $\mu$s \\
    $T_2^\mathrm{echo}$ & 55 $\pm$ 13 $\mu$s & 34 $\pm$ 6 $\mu$s \\
    \hline
    \hline
  \end{tabular}
  \caption{Measured qubit and resonator parameters.
  $\omega_\mathrm{R}$ and $\kappa$ are the bare resonator frequency and coupling rate to the transmission line.
  $\omega_{01}$ and $\alpha$ are the qubit transition frequency and anharmonicity.
  $g$ is the coupling rate between the resonator and the qubit.
  $\chi$ is the dispersive shift such that $\omega_\mathrm{R} \pm \chi$ are the dressed frequencies of the readout resonator.
  $\Gamma_{Purcell}$ is the Purcell rate \cite{sete_quantum_2015}.
  $T_1$ and $T_2^\mathrm{echo}$ are the energy-relaxation time and decoherence time of the qubit: nominal values and uncertainties are the median and interquartile range, respectively, of $\approx 2.5$ thousand measurements spanning $2.5$ days.
  }
  \label{tab:parameters}
\end{table}

\begin{figure*}
  \centering
  \includegraphics{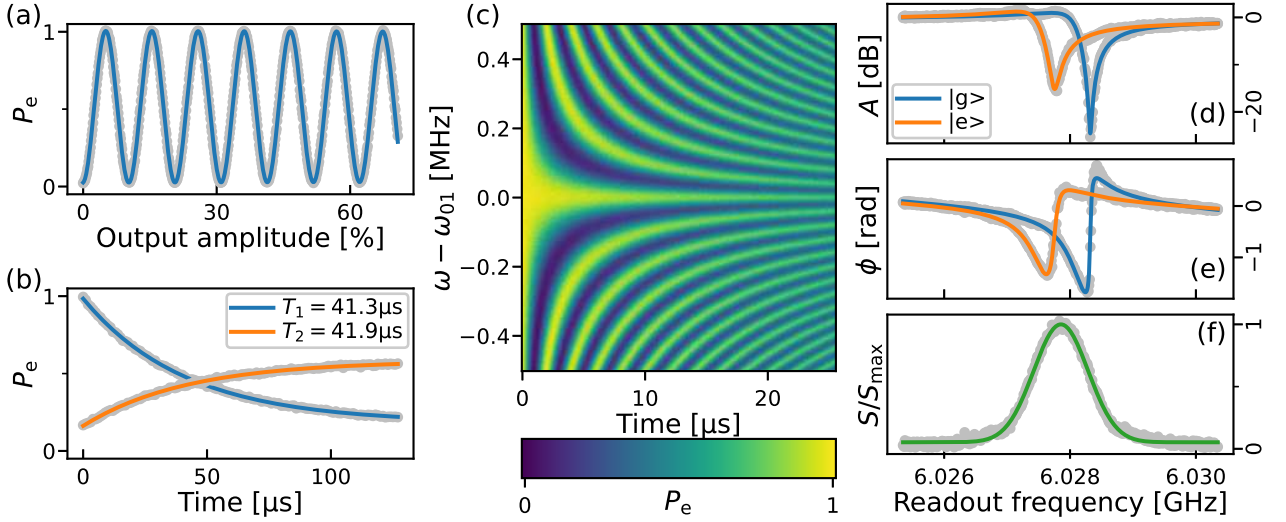}
  \caption{Characterization of qubit 2 with Presto.
  Gray dots show measurement data, solid lines show fits to theory.
  (a) Rabi oscillations with a sequence of ten DRAG-compensated control pulses, each with a sine-squared shape and duration 20~ns.
  The fitted amplitude for a $\pi$ gate is 51.86\% of the DAC full-scale range, or -5.9~dBm.
  (b) Measurements of energy-relaxation time $T_1$ and decoherence time $T_2$ from a Ramsey-echo experiment.
  (c) Chevron pattern from a series of Ramsey measurements.
  (d, e) Pulsed resonator spectroscopy showing the steady-state resonator response $R_{\mathrm{g/e}}(f)$ when the qubit is prepared in the ground (blue) and excited (orange) states: (d) amplitude $A$; (e) phase $\phi$.
  (f) Separation in phase space between the two steady-state responses: $S(f) = |R_\mathrm{e}(f) - R_\mathrm{g}(f)|$.
  The frequency of maximum separation is the optimal readout frequency.
  }
  \label{fig:combo_meas}
\end{figure*}

We showcase Presto's capabilities on a sample consisting of two fixed-frequency superconducting qubits coupled by a frequency-tunable coupler \cite{mckay_universal_2016}.
A description of the measurement setup is given in Appendix~\ref{apx:setup}, including details of the bandpass filters which are necessary with DDS at microwave frequencies.
The qubit circuit is described in detail in Ref.~\citenum{bengtsson2020improved}.
Each qubit is controlled by a dedicated microwave line, and each is dispersively coupled to its own readout resonator.
Both resonators are coupled in a notch configuration to a single microwave line for frequency-multiplexed readout.
Table~\ref{tab:parameters} shows a summary of the measured qubit and resonator parameters.
These parameters are obtained with standard single-qubit characterization techniques, as shown in Fig.~\ref{fig:combo_meas}.

To control the qubits we use DRAG-compensated \cite{KrantzP2019Aqeg} microwave pulses with a $\sin^2(x)$ envelope, where $x=\pi t / \tau$ and $\tau = 20~\mathrm{ns}$ is the duration of one pulse.
To readout the qubit, we apply a $1.4~\mathrm{\mu s}$-long pulse composed of four segments with constant complex amplitude.
This readout pulse is an adaptation of the CLEAR pulse \cite{McClure2016clear}: the first two segments rapidly populate the readout resonator with photons, and the last two segments reset the resonator to its ground state, regardless of the measured state of the qubit.
The maximum instantaneous power in the readout pulse is approximately $-115$~dBm at the sample input.

The code used to perform the measurements reported here is released as open source under the MIT license, and is available online in a repository hosted on GitHub \cite{presto-measure}.

\subsection{Active qubit reset}
\label{sec:reset}

\begin{figure}
  \centering
  \includegraphics{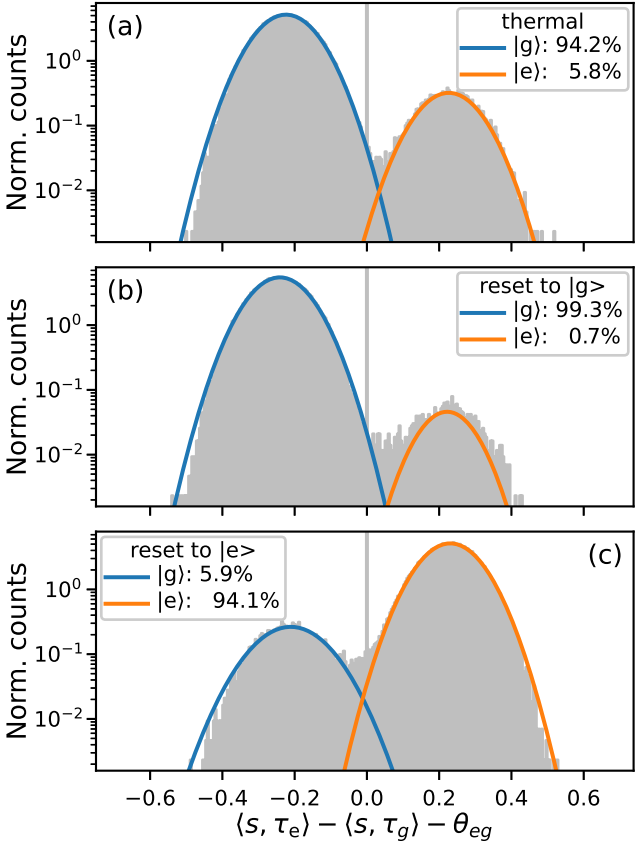}
  \caption{Histograms of single-shot readouts of the qubit in thermal equilibrium (a), and after a conditional pulse to reset the qubit in its ground (b) or excited (c) state.
  The histograms (gray) are calculated over $10^5$ measurements with 316 bins.
  The resulting distribution is fitted to a bimodal Gaussian distribution, and the individual fitted components are plotted in blue for the ground state and orange for the excited state.
  }
  \label{fig:reset}
\end{figure}

We use the template-matching and feedback features of Presto to perform single-shot readout of the state of qubit 2, actively resetting the qubit to the ground state if we measure it in the excited state (Fig.~\ref{fig:reset}).
The histogram in Fig.~\ref{fig:reset}(a) is built from $10^5$ independent measurements of the qubit state starting from thermal equilibrium, achieved by waiting a time $500~\mathrm{\mu s} \gg T_1$ between each measurement.
We perform template matching over a 1~$\mu$s-long window, projecting in real time each measured single-shot trace $s$ onto two calibrated templates $\tau_\mathrm{g}$ and $\tau_\mathrm{e}$ (uploaded to Presto before the measurement) representing the average ground- and excited-state responses, respectively.
As described in Sec.~\ref{sec:feedback}, Presto's feedback engine compares the template-matching outcome to a threshold, such that $\left< s, \tau_\mathrm{e} \right> - \left< s, \tau_{g} \right> \ge \theta_{eg}$ corresponds to the excited state being the most probable outcome of the measurement.
Based on this comparison we enable or disable the output of a $\pi$ pulse to flip the state of the qubit.
A second readout and template-matching operation is then performed immediately after this conditional reset to assess its effectiveness.

From the first of the two readout pulses we can determine the excited-state population $P_\mathrm{therm}$ when the qubit is in equilibrium with its environment.
Fitting a bimodal Gaussian distribution to the measurements we obtain an excited-state population of $P_\mathrm{therm}=5.8\%$, roughly equivalent to an effective temperature of $T_\mathrm{eff} = \hbar \omega_{01} / [k_\mathrm{B} \log (1/P_\mathrm{therm} - 1)] = 71~\mathrm{mK}$, significantly higher than the temperature measured by the cryostat thermometer at $10~\mathrm{mK}$.

Nevertheless, the fact that the quantum circuit is in a mixed state when quiescent, provides an opportunity to test Presto's feedback engine.
Using the same fitting procedure on the data from the second readout (after the conditional reset pulse),
we find an excited-state population of 0.7\% [see Fig.~\ref{fig:reset}(b)], roughly equivalent to an effective temperature of 40~mK.
Presto's feedback engine is cooling the qubit.

Preparing the qubit in $\left| \mathrm{e} \right>$ with one round of feedback results in an excited-state population of 94.2\% [See Fig.~\ref{fig:reset}(c)].
The lower state-preparation fidelity compared with $\left| \mathrm{g} \right>$ is consistent with qubit decay during the feedback iteration.

From the fit to the histogram of Fig.~\ref{fig:reset}(a), we also obtain an indication of the measurement infidelity due to overlap errors.
We analyze the bimodal distribution of template matching results, obtaining the mean $\mu_\mathrm{g}<0$ and standard deviation $\sigma_\mathrm{g}$ for the ground state, and $\mu_\mathrm{e}>0$ and $\sigma_\mathrm{e}$ for the excited state.
The overlap error is then $\epsilon_\mathrm{overlap}=1-(\epsilon_\mathrm{g} + \epsilon_\mathrm{e})/2$, where $\epsilon_i$ is the probability of assigning the wrong state given a qubit in state $i$: $\epsilon_i = [1-\mathrm{erf}(x_i)]/2$, where $\mathrm{erf}$ is the error function and $x_i=|\mu_i| / (\sqrt{2} \sigma_i)$.
In our measurement $\epsilon_\mathrm{overlap} = 9.7 \times 10^{-4}$, which imposes an upper limit on measurement fidelity of $\mathcal{F}_\mathrm{meas}<1-\epsilon_\mathrm{overlap}=99.903\%$.

\subsection{Randomized benchmarking}

\begin{figure}
  \centering
  \includegraphics{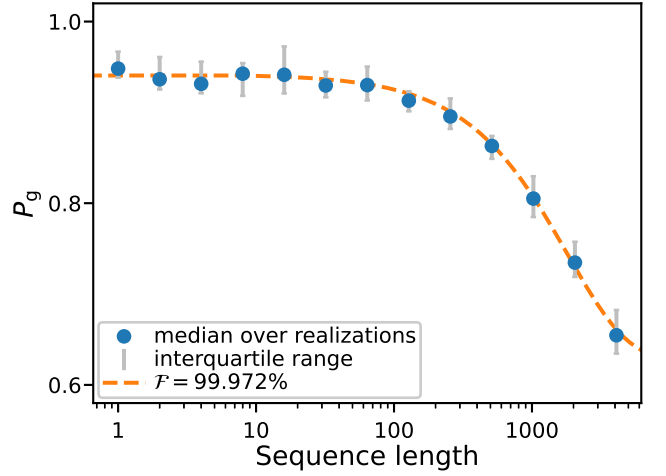}
  \caption{Single-qubit randomized benchmarking, showing the probability of measuring the qubit in the ground state as function of the number of Clifford gates in the sequence.
  For each sequence length $m$, 50 different realizations of the random sequence are applied.
  For each realization, the measurement is repeated and averaged 1000 times.
  Blue dots and gray bars show the median and interquartile range, respectively, over the single realizations.
  An exponential decay of the form $A \alpha ^ m + B$ is fit (orange dashed line) to the average response over all realizations (blue dots).
  The obtained error per Clifford is $(1-\alpha)/2 = (2.8 \pm 1.2) \times 10^{-4}$, corresponding to an average fidelity of $\mathcal{F} = 99.972 \pm 0.012 \%$.
  }
  \label{fig:rb}
\end{figure}

We characterize the quality of the control signals from Presto with randomized benchmarking on qubit 2.
We use Qiskit \cite{Qiskit} to generate random sequences of Clifford gates.
Each sequence is trans-compiled into a series of $\sqrt{X}$ gates ($\pi/2$ pulses) and virtual $Z$ gates (phase rotations of all subsequent $\pi/2$ pulses) \cite{McKayDavidC.2017EZgf}.
We execute 50 different random sequences of gates starting from thermal equilibrium, averaging over 1000 measurements for each sequence.
The results are shown in Fig.~\ref{fig:rb} for 13 different sequence lengths.
We obtain an average single-qubit fidelity of $99.972 \pm 0.012\%$, as described in the caption to Fig.~\ref{fig:rb}.
This fidelity is consistent with the theoretical limit set by the qubits' $T_1$ and $T_2$ times \cite{abad2021rbfidelity} $\mathcal{F}_{\sigma_x} = 1 - \frac{\Gamma_1 + \Gamma_\phi}{3}\tau = 99.971 \pm 0.004 \%$, where $\Gamma_\phi = \Gamma_2 - \Gamma_1 / 2$ is the pure dephasing rate, $\Gamma_i = 1/T_i$ and $\tau=20~\mathrm{ns}$ is the duration of a control pulse.
Thus, at this level of fidelity, we conclude that Presto is an insignificant source of error.

\subsection{iSWAP}

\begin{figure}
  \centering
  \includegraphics{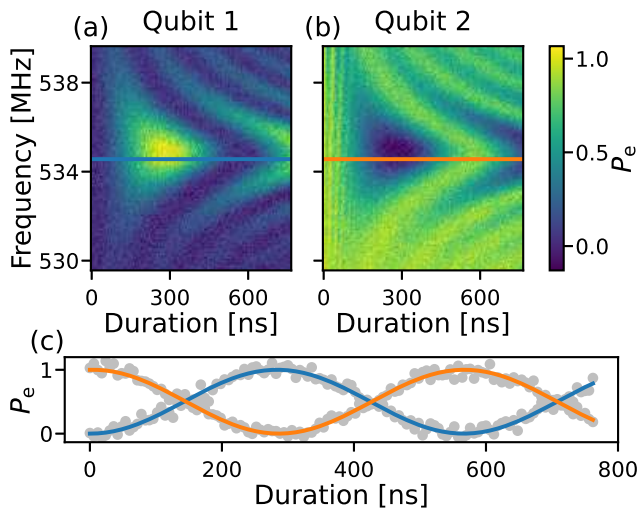}
  \caption{Tune-up of an iSWAP gate.
  The top panels show the probability $P_\mathrm{e}$ of measuring qubit 1 (a) and qubit 2 (b) in the excited state, measured with a frequency-multiplexed readout.
  $P_\mathrm{e}$ is shown as function of the duration and frequency of a pulse on the tunable coupler.
  The lower panel (c) shows a line cut of $P_\mathrm{e}$ along the blue and orange lines: gray dots are measured data, solid lines are numerical fits to a cosine function.
  }
  \label{fig:iswap}
\end{figure}

Figure~\ref{fig:iswap} shows the calibration and characterization of an exchange-type two-qubit gate \cite{ganzhorn2020benchmarking} (iSWAP).
A DC flux of $0.26~\Phi_0$ is applied to the coupler to select its operating point for the entire measurement sequence.
We start the sequence by applying a $\pi$ pulse to qubit 2, to initialize the state $\left|01\right>$, \textit{i.e.} qubit 1 in the ground state and qubit 2 in the excited state.
We then apply a flux drive with $0.21~\Phi_0$ amplitude to the tunable coupler, varying the drive frequency and duration to enable the exchange of quantum information between the two qubits.
At the end of this coupler pulse we simultaneously measure the state of both qubits with a frequency-multiplexed readout pulse.
Figure~\ref{fig:iswap} shows both qubit measurement probabilities as a function of the frequency and duration of the iSWAP pulse.
A cut of this data set at 534.5~MHz shows that we can perform the iSWAP operation in 300~ns.
Further experiments are needed to optimize this two-qubit gate.
A future publication will discuss more thoroughly the advantages that Presto's phase synchronization and DDS offer for two-qubit gates, as well as the achievable gate fidelities.

\section{Conclusions}

We described Presto, a general-purpose microwave platform for qubit readout and control based on the third-generation RFSoC.
Presto uses digital methods that enable microwave-signal synthesis and real-time analysis, without external local oscillators and analog IQ mixers for up- and down-conversion.
With 16 input and 16 output channels in a 2U 19-inch rack format, Presto offers high channel density with all channels synchronized by one master clock.
Multiple Presto units can be synchronized with deterministic latency, providing the large number of synchronous channels needed for the control of future quantum processors.

A Python API provides flexible and high-level access to the FPGA and other digital hardware, required in a general-purpose instrument for experimentation with superconducting quantum circuits.
The digital nature of this microwave platform, with tight integration between data converters, FPGA and processing system, makes it possible to easily implement a variety of features through programming.
Combined with its general-purpose analog front-end, Presto is easily adapted with additional features to meet the demands of future experiments.

We showed how Presto is used to characterize superconducting qubits.
We used its template-matching feature for single-shot state discrimination and its low-latency feedback to implement conditional reset of a single qubit.
We showed randomized benchmarking of single-qubit gates, demonstrating a fidelity of 99.972\%, limited by the qubit's relaxation and dephasing times.
We also implemented an iSWAP two-qubit gate.
In all of these experiments, Presto was the only instrument used for control and readout.

These initial results are encouraging and they demonstrate that Presto and the DDS techniques upon which it is built, are not an additional source of error at this level of fidelity.
A next step in testing Presto is randomized benchmarking of two qubit gates.
Beyond that we look forward to using Presto for the control of more coherent qubits, and toward demonstrating Presto's synchronous multi-channel capability on the control of multi-qubit chips.

\begin{acknowledgments}
The authors thank Joe Aumentado and the Advanced Microwave Photonics Group at NIST Boulder, CO for providing the Josephson parametric amplifier used in this study.
The authors acknowledge funding from the Knut and Alice Wallenberg Foundation (KAW) through the Wallenberg Centre for Quantum Technology (WACQT) and the Wallenberg Launch Pad (WALP).
MOT, RB and DBH are part owners of Intermodulation Products AB, which manufactures and sells the microwave platform described in this manuscript.
\end{acknowledgments}

\appendix

\section{Continuous-wave measurements}
\label{apx:cw}

\begin{figure}
  \centering
  \includegraphics{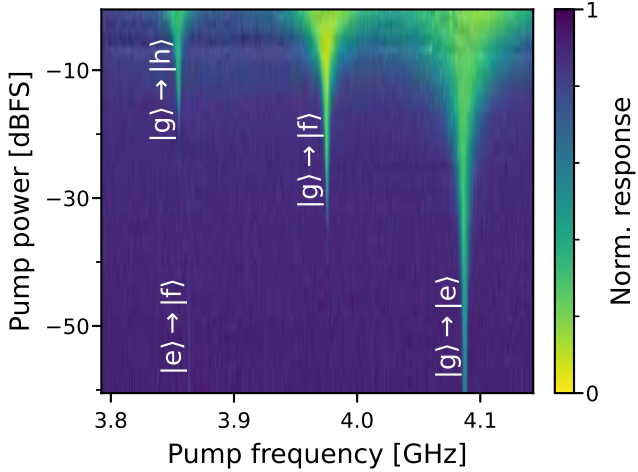}
  \caption{
    Tow-tone spectroscopy in continuous-wave mode.
    The image shows the response of the resonator to a probe tone with constant amplitude at 6.03~GHz,
    while a pump tone is swept in frequency and power.
    The white text marks the visible transition frequencies.
    Transitions $\left|\mathrm{g}\right> \rightarrow \left|\mathrm{f}\right>$ and $\left|\mathrm{g}\right> \rightarrow \left|\mathrm{h}\right>$ are two-photon and three-photon transitions, respectively.
  }
  \label{fig:twotone_cw}
\end{figure}

The continuous-wave mode of operation turns Presto into a multifrequency lock-in amplifier \cite{mla}.
Presto generates and outputs a microwave frequency comb with up to 192 tones, with frequency, amplitude and phase at each tone independently defined by the user.
As in the pulsed mode described in the main text, the comb is initially generated at sampling rate 1~GS/s and then interpolated up to the sampling rate of the DAC.
A digital IQ mixer and NCO up-convert the comb to the desired passband, in a 1~GHz band centered around a 0-9~GHz carrier.
On the input side, 192 demodulators are available to measure the lock-in components of an incoming signal, all relative to one common reference phase.
The frequency and phase of the demodulators are independent from that of the generators for the output comb.
The demodulation rate is selected by the user between 1~Hz and 1~MHz.

Figure~\ref{fig:twotone_cw} shows a two-tone spectroscopy measurement using the continuous-wave mode.
The spectral line on the right is the $\left|\mathrm{g}\right> \rightarrow \left|\mathrm{e}\right>$ transition,
the faint line on the bottom left is the $\left|\mathrm{e}\right> \rightarrow \left|\mathrm{f}\right>$ transition,
the line at the center is the two-photon $\left|\mathrm{g}\right> \rightarrow \left|\mathrm{f}\right>$ transition,
and the line on the upper left is the three-photon $\left|\mathrm{g}\right> \rightarrow \left|\mathrm{h}\right>$ transition.

\section{Calibration of reference templates for readout}
\label{apx:match}

\begin{figure}
  \centering
  \includegraphics{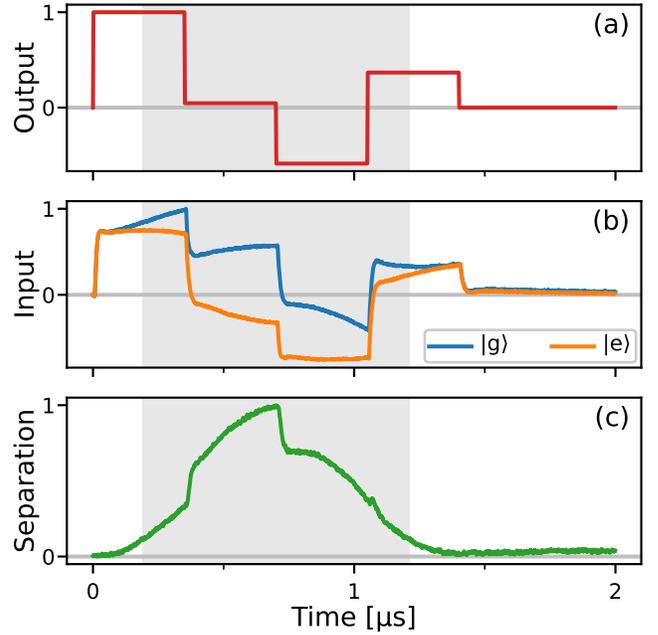}
  \caption{
    Calibration of template matching for qubit readout.
    (a) Normalized I quadrature of the readout pulse with four segments of constant amplitude and phase.
    (b) Normalized I quadrature of the measured resonator response when the qubit is in its ground (blue) and excited (orange) state.
    (c) Normalized distance in the complex plane between the traces in panel (b).
    The gray area shows the template-matching window that maximizes the measurement fidelity.
  }
  \label{fig:AppendixB}
\end{figure}

We design the readout pulse to maximize the measurement fidelity and minimize the number of photons left in the resonator after the pulse.
We use a pulse composed of four constant-amplitude segments \cite{McClure2016clear}, each of duration $350~\mathrm{ns} \approx 1/\kappa$ [Fig.~\ref{fig:AppendixB}(a)].
Two ring-up segments rapidly increase the number of photons in the cavity, thus increasing the state separation and enhancing the readout fidelity \cite{sete_quantum_2015}.
Two ring-down segments rapidly force the resonator back to its ground state at the end of the pulse.
We optimize the amplitude and phase of the four segments by solving the linear differential equation that models the resonator for the parameters given in Table~\ref{tab:parameters}.

With this readout pulse we measure the response of the resonator when the qubit is in the ground and excited state.
We first acquire two preliminary reference templates $\tilde{\tau}_\mathrm{g}$ and $\tilde{\tau}_\mathrm{e}$ by preparing the qubit in its ground state and excited state (applying a $\pi$~pulse), respectively.
Each template is formed by averaging the resonator response to $10^6$ readout pulses.
Due to the rather large thermal population [Fig.~\ref{fig:reset}(a)], $\tilde{\tau}_\mathrm{g}$ is contaminated by a small fraction of resonator response with the qubit in its excited state, and vice versa for $\tilde{\tau}_\mathrm{e}$ .

To further refine the reference templates, we perform a second measurement with two readout pulses separated by a small delay.
We perform template matching (Sec.~\ref{sec:acquisition}) on the first readout pulse with $\tilde{\tau}_\mathrm{g}$ and $\tilde{\tau}_\mathrm{e}$, while storing the response of the resonator to the second readout.
We then post-select the measured traces from the second readout, conditioned on the outcome of template matching on the first readout.
Thus we average only the traces with the qubit truly in its ground (excited) state to obtain refined reference templates $\tau_\mathrm{g}$ ($\tau_\mathrm{e}$).
These refined reference templates are shown in Fig.~\ref{fig:AppendixB}(b).

Finally, we select the optimal time window for template matching, which is limited to 1022~ns in the current firmware.
We select the 1022-ns window that maximizes the observed state separation $d = \int \left| \tau_\mathrm{e} - \tau_\mathrm{g} \right| \mathrm{d}t$.
Note that the reference templates $\tau_\mathrm{g}$ and $\tau_\mathrm{e}$ are complex valued and $d$ is the integral of their separation in the complex plane.
The time window that maximizes $d$ is the optimal template-matching window for performing state discrimination, since measurement fidelity increases with state separation \cite{sete_quantum_2015}.
Figure~\ref{fig:AppendixB}(c) shows the measured state separation $d$ and the time window for optimal matching.

\section{Active qutrit reset}
\label{apx:qutrit}

Presto's feedback mechanism can be used to actively reset a quantum 3-level system, or qutrit \cite{Magnard2018}.
The procedure is analog to the active qubit reset described in Sec.~\ref{sec:feedback} and Sec.~\ref{sec:reset}.
Say we have calibrated reference templates for the qutrit in its ground $\tau_\mathrm{g}$, first excited $\tau_\mathrm{e}$ and second excited $\tau_\mathrm{f}$ states.
We compute the thresholds $\theta_{ij} = (\lVert \tau_i \rVert^2 - \lVert \tau_j \rVert^2)/2$,
where $\lVert \tau \rVert^2 = \left< \tau, \tau \right>$ and $i,j \in \left\{\mathrm{e},\mathrm{g},\mathrm{f}\right\}$.
For convenience, we define the comparison result $R_{ij} \stackrel{\text{def}}{=} \left[ \left< s, \tau_\mathrm{i} \right> - \left< s, \tau_{j} \right> \ge \theta_{ij} \right]$, which can be true or false (1 or 0).

We program the first template-matching pair with $\tau_\mathrm{e}$ and $-\tau_\mathrm{g}$ and a threshold of $\theta_\mathrm{eg}$.
This comparison will yield $R_\mathrm{eg}=1$ if the measured trace matches $\tau_\mathrm{e}$ better than $\tau_\mathrm{g}$, and 0 otherwise.
A second template-matching pair is programmed with $\tau_\mathrm{f}$ and $-\tau_\mathrm{e}$ and a threshold of $\theta_\mathrm{fe}$.
The third pair is programmed with $\tau_\mathrm{g}$, $-\tau_\mathrm{f}$ and $\theta_\mathrm{gf}$.

We program two possible output pulses, $\pi_\mathrm{eg}$ and $\pi_\mathrm{fg}$, that reset the qutrit to its ground state, conditioned on it being in the first or second excited state, respectively.
The reset from f to g could be implemented as the sequence of two pulses, intermediately putting the qutrit in its first excited state.
The output of $\pi_\mathrm{eg}$ should be enabled if $R_\mathrm{eg}=1$ and $R_\mathrm{fe}=0$, and disabled otherwise.
Notice the value yielded by the third comparison $R_\mathrm{gf}$ is irrelevant in this case.
Similarly, $\pi_\mathrm{fg}$ should be output if $R_\mathrm{fe}=1$ and $R_\mathrm{gf}=0$.
For a qutrit in its ground state, $R_\mathrm{eg}=0$ and $R_\mathrm{gf}=1$, and no reset pulse should be output.
The values yielded by the comparisons for the three possible input states and the desired output pulse are summarized in Table~\ref{tab:qutrit}.

\begin{table}[h]
  \begin{center}
    \begin{tabular}{c || c | c | c || c}
      \hline
      \hline
      Measured state & $R_\mathrm{eg}$ & $R_\mathrm{fe}$ & $R_\mathrm{gf}$ & Output pulse \\
      \hline
      $\left|\mathrm{g}\right>$ & 0 & --- & 1 & --- \\
      $\left|\mathrm{e}\right>$ & 1 & 0 & --- & $\pi_\mathrm{eg}$ \\
      $\left|\mathrm{f}\right>$ & --- & 1 & 0 & $\pi_\mathrm{fg}$ \\
      \hline
      \hline
    \end{tabular}
  \end{center}
  \caption{Truth table for comparison operators in a qutrit readout.}
  \label{tab:qutrit}
\end{table}

The Boolean operator in the feedback engine encodes this truth table with three bits of input (the result of the three comparisons) and two bits of output (the enable signals for the reset pulses).
In the event sequencer the user schedules a template matching operation on the three template-matching pairs, followed by the release of both pulses $\pi_\mathrm{eg}$ and $\pi_\mathrm{fg}$, immediately after the readout.
The two output bits from the Boolean operator control which pulse is sent to the qutrit.

\section{Measurement setup}
\label{apx:setup}

\begin{figure}
  \centering
  \includegraphics{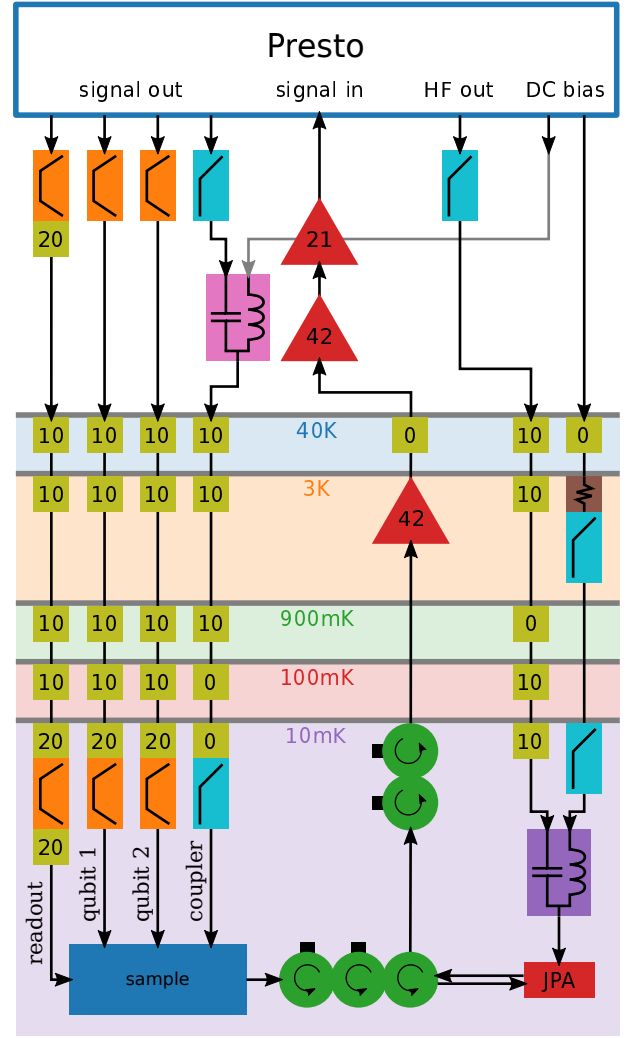}
  \caption{Schematic drawing of the room-temperature and cryogenic setup.
  An 8-channel Presto is used to generate all the readout and control signals to the two-qubit quantum processor,
  including DC biases and the pump to the JPA.
  }
  \label{fig:cryo_setup}
\end{figure}

Figure~\ref{fig:cryo_setup} shows a schematic drawing of the room-temperature and cryogenic setup.
We use microwave filters from Mini-Circuits, both at room temperature and inside the dilution refrigerator, to suppress unwanted image signals generated by the high-speed DACs and to limit the noise bandwidth.
There are band-pass filters on the readout line (VBFZ-6260-S+) and on the two qubit-control lines (VBFZ-4000-S+), and low-pass filters on the coupler line (VLFX-650+).
A bias tee (Mini-Circuits ZFBT-4R2GW+) at room temperature combines the AC drive to the coupler with a DC bias.
This configuration for the coupler line is not optimal and is likely the main contribution to the high thermal population observed in the experiments.

The readout line contains a cascade of amplifiers.
An InP HEMT amplifer is mounted at 3~K (Low Noise Factory LNF-LNC4\_8C) followed by two room temperature amplifiers (Low Noise Factory LNF-LNR4\_8ART and Mini-Circuits ZX60-83LN-S+) giving a total gain of about 105~dB at 6~GHz.
For single-shot readout we also use a JPA provided by NIST Boulder, CO \cite{lecocq2017jpa}.
The JPA is pumped by Presto: one of the two high-frequency clock outputs creates the pump at twice the readout frequency ($\approx 12~\mathrm{GHz}$), and one of the DC-bias outputs controls the operating point of the JPA.
The RF signal is low-pass filtered (Mini-Circuits ZLSS-14G-S+) to remove unwanted harmonics and the DC line has a 1~k$\Omega$ bias resistor at 3~K followed by two low-pass filters (Mini-Circuits SLP-1.9+).
The RF and DC signals are combined at the mixing chamber with a diplexer (Mini-Circuits ZDSS-3G4G-S+).

\section{Abbreviations}
\label{apx:abbrev}

Here follows a list of abbreviations used in this manuscript and their description.

\begin{description}
\item [5G] fifth-generation technology standard for broadband cellular networks
\item [AC] alternating-current
\item [ADC] analog-to-digital converter
\item [API] application-programming interface
\item [AWG] arbitrary-waveform generator
\item [CPU] central-processing unit
\item [CMOS] complementary metal-oxide-semiconductor
\item [CLEAR] cavity-level excitation-and-reset
\item [DAC] digital-to-analog converter
\item [DC] direct-current
\item [DDS] direct digital synthesis
\item [DRAG] derivative reduction by adiabatic gate
\item [DSP] digital signal processing
\item [FIFO] first-in first-out (buffer)
\item [FPGA] field-programmable gate array
\item [HEMT] high-electron-mobility transistor
\item [InP] indium phosphide
\item [JPA] Josephson parametric amplifier
\item [LVTTL] low-voltage transistor-transistor logic
\item [MIT] Massachusetts Institute of Technology
\item [MTS] multi-tile synchronization
\item [NCO] numerically-controlled oscillator
\item [NIST] National Institute of Standards and Technology
\item [PCB] printed circuit board
\item [PLL] phase-locked loop
\item [RF] radio-frequency
\item [RFSoC] radio-frequency system on a chip
\item [RPU] real-time processing unit
\item [SDR] software-defined radio
\item [SDRAM] synchronous dynamic random-access memory
\item [TWPA] travelling-wave parametric amplifier
\end{description}

\nocite{*}
\bibliography{mybib}

\end{document}